\documentclass{ifacconf}

\usepackage{graphicx}      % include this line if your document contains figures
\usepackage{natbib}        % required for bibliography
\usepackage{amsmath}
\usepackage{amssymb}
\usepackage{algorithm}
\usepackage{color}
\usepackage{arydshln} % in the preamble
\usepackage{algorithmic}
\usepackage{bm}

\usepackage{dsfont}
\usepackage{enumitem}
\usepackage{balance}

\def\1{\mathds{1}}

\def\R{\mathbb{R}}

\newcommand{\mc}{\mathcal}
\usepackage{url}

\usepackage{tcolorbox}
\newcounter{myblockcounter}
\renewcommand{\themyblockcounter}{BLK-\arabic{myblockcounter}}

\newtcolorbox[use counter=myblockcounter]{mybox}[2]{%
  colback=gray!20, 
  colframe=gray!50, 
  sharp corners=south, 
  title=Block \themyblockcounter: #1, 
  label=#2
}

\usepackage[prependcaption,colorinlistoftodos]{todonotes}

\begin{document}
\begin{frontmatter}

\title{Free Parametrization of $\mathcal{L}_2$-Bounded Structured State-Space Controllers for Nonlinear Control with Stability Guarantees\thanksref{footnoteinfo}}

\thanks[footnoteinfo]{This work is funded by the Swiss National Science Foundation (grant no. 200021-204962), NCCR Automation, a National Centre of Competence in Research, funded by the Swiss National Science Foundation (grant no. 51NF40\_225155), and the NECON project (grant no. 200021-219431).
$\dagger$The authors contributed equally. Corresponding authors: \texttt{muhammad.zakwan@inspire.ch}, and \texttt{l.massai@epfl.ch}}

\author[First]{Muhammad Zakwan$\dagger$} 
\author[Second]{Leonardo Massai$\dagger$} 
\author[First]{Efe C. Balta}
\author[Second]{Giancarlo Ferrari-Trecate}

\address[First]{Control \& Automation Group, Inspire AG, 8005 Zürich, Switzerland \& with Automatic Control Laboratory (IfA), ETH Zürich, 8092 Zürich, Switzerland.}
\address[Second]{Laboratoire d’Automatique, EPFL, 1015 Lausanne, Switzerland.}

\begin{abstract}                % Abstract of 50--100 words
%Designing stabilizing controllers for nonlinear systems while optimizing complex cost functions remains a formidable challenge. Neural Networks (NNs), renowned for their expressive power, offer a promising approach to parametrize control policies that achieve high performance. However, their inherent sensitivity to small input perturbations can destabilize the closed-loop system, undermining reliability. Many existing methods address this by imposing constraints on the controller's parameter space to ensure stability, but these constraints often lead to computationally intensive optimization processes.
%To overcome these limitations, we propose leveraging Structured State-Space Models (SSMs) to design discrete-time control policies for nonlinear systems. Our approach guarantees closed-loop stability and finite $\mathcal{L}_2$ gains, independent of the controller's optimization parameters. SSMs have recently emerged as a powerful alternative due to their ability to capture long-range dependencies with linear complexity relative to sequence length---unlike the quadratic complexity of transformer-based architectures. This eliminates the need for parameter constraints during optimization, enabling the use of efficient, standard techniques such as gradient-based methods.
%We demonstrate the effectiveness of the proposed approach through a case study on formation control of mobile robots, where the controllers ensure collision and obstacle avoidance while maintaining stability and performance.
Designing stabilizing control policies for nonlinear systems while optimizing complex objectives remains a formidable challenge. 
Neural Networks (NNs), despite their expressive power, can be highly sensitive to small input perturbations and can easily destabilize the closed-loop system. 
Existing approaches often impose explicit constraints on the controller’s parameters to ensure stability, but this typically leads to extra computational overhead.
To address this issue, we leverage recently proposed Structured State-Space Models (SSMs) to parametrize discrete-time control policies for nonlinear systems. Our key contribution is a new free parametrization of Linear Time Invariant (LTI) systems with a prescribed $\mathcal{L}_2$-gain, which we use to construct the L2-Recurrent Unit (L2RU)  architecture, an SSM layer that enforces the desired $\mathcal{L}_2$-bound \emph{by design}. This result can be leveraged to guarantee closed-loop stability via the small-gain theorem or the so-called performance-boosting framework, independently of the controller’s optimization parameters, thereby enabling fully unconstrained optimization of general nonlinear objectives. Furthermore, the structure induced by the proposed parametrization allows efficient processing of long input sequences, as it is highly parallelizable through algorithms such as parallel scan.
We demonstrate the effectiveness of this approach on a formation control task for mobile robots, where the L2RU-based controller ensures collision and obstacle avoidance while maintaining stability and performance.

\end{abstract}


\end{frontmatter}
%===============================================================================

\section{Introduction}

In recent years, there has been a surge of research interest in deep-learning architectures for control. 
A wide range of increasingly sophisticated models, from Recurrent Neural Networks (RNNs)~\citep{bonassi_recurrent_2022} to Transformers \citep{vaswani_attention_2023}, have been proposed for nonlinear system identification and optimal control, where they serve as expressive parametrizations for high-performance controllers. 
A central challenge in these applications is the necessity to guarantee stability and robustness of the closed-loop system while optimizing complex, nonlinear objectives. Classical control design explicitly enforces stabilizing constraints, but neural network architectures often lack built-in robustness: small input perturbations may propagate unpredictably and destabilize the closed-loop system~\citep{zakwan_robust_2023, zakwan2024neural}. 

In this paper, we are interested in designing output feedback controllers for dynamical systems that minimize a highly nonlinear cost while simultaneously ensuring stability of the closed-loop system. 
In general, designing a stable closed-loop system with NNs is challenging because the requirement for stability imposes strict constraints on the controller’s parameters. 
This often results in nonconvex optimization problems that are difficult to solve.

A classical approach to ensuring tractability is to leverage the small-gain theorem~\citep{van_der_schaft_small-gain_2017}.
Specifically, if the plant has a known (or estimated) $\mathcal{L}_2$-gain $\hat{\gamma}$ and the controller has an $\mathcal{L}_2$-gain $\gamma$, then the feedback interconnection is guaranteed to be $\mathcal{L}_2$-stable provided that $\gamma \hat{\gamma} < 1$. Another powerful method is the so-called performance-boosting framework, which parametrizes all and only stability-preserving controllers of a given pre-stabilized plant but requires an explicit $\mc L_2$-bound for the controller in the presence of model mismatch (see \cite{galimberti_parametrizations_2025} for references).
However, enforcing an $\mathcal{L}_2$-gain constraint during learning for general dynamical controllers remains nontrivial.
The key idea of this paper is to overcome this difficulty by parametrizing controllers as Structured State-Space Models (SSMs)~\citep{gu_mamba:_2024} with a prescribed $\mathcal{L}_2$-gain, using the architecture introduced in~\cite{massai_free_2025} under the name \emph{$\mathcal{L}_2$-bounded Linear Recurrent Unit} (L2RU).
SSMs are composed of several modular blocks, each comprising a linear recurrent unit surrounded by nonlinear static functions that form the so-called \emph{scaffolding} of the SSM.
Recently, SSMs have gained attention in sequence modeling due to their linear computational complexity with respect to sequence length, in contrast to the quadratic complexity of Transformer-based architectures.
While the recurrent units in SSMs are stable by design, their use as control policies does not inherently guarantee closed-loop stability.
In this work, we introduce a novel free parametrization of the SSM-based L2RU architecture that generalizes the one in~\cite{massai_free_2025} and offers improved computational efficiency. Crucially, the parametrization is free in the sense that for every value of the parameters, the associated L2RU is guaranteed to have a prescribed $\mathcal{L}_2$-gain $\gamma$.
As a consequence, parametrizing the controller with the L2RU unlocks the possibility to decouple stability from optimization: by choosing $\gamma$ to satisfy the small-gain condition with the plant, closed-loop stability is ensured by design, while the controller parameters can be tuned via fully unconstrained optimization of arbitrary nonlinear objectives. 
Beyond the small-gain condition, the $\mathcal{L}_2$-bounded nature of the L2RU also makes it a natural choice for parametrizing highly nonlinear controllers within the aforementioned performance-boosting framework ~\citep{galimberti_parametrizations_2025}.

{\bf Related Work:}
Several works have explored the use of NN-parametrized control policies for dynamical systems, ensuring closed-loop stability \emph{by design}, allowing one to bypass computationally expensive \emph{a posteriori} verification routines and projections during optimization.
For example, internal model control~\citep{furieri2024learning, distr} have been employed to design controllers for nonlinear systems.
Additionally, contraction theory~\citep{FB-CTDS} has been applied for set-point stabilization of control-affine systems~\citep{zakwan2024neuralcontrol}.
However, these approaches are computationally expensive to train compared to SSMs, which leverage parallel scan for efficient inference \citep{blelloch_prefix_2004}.

A stream of works, such as~\cite{furieri_distributed_2022, zakwan2024neural, zakwan2024neuralporthamiltonianmodelsnonlinear}, focuses on port-Hamiltonian models to parametrize nonlinear control policies that guarantee closed-loop stability while minimizing a general cost.
Key differences arise in their use of continuous-time control policies, which are challenging to implement in real-world applications.
Specifically, discretization of these policies does not preserve the $\mathcal{L}_2$-gain, leading to a loss of stability guarantees.
While~\cite{zakwan2024neuralporthamiltonianmodelsnonlinear} proposes to use discrete gradients for discretization that preserve the $\mathcal{L}_2$-gain, the resulting discrete-time forward equation becomes \emph{implicit}, hence, making the implementation cumbersome.
Furthermore, these methods rely on Neural Ordinary Differential Equations (NODEs)~\citep{chen2018neural}, which require numerical integration and are computationally expensive compared to the fast parallel scan-based forward inference of SSMs.

In the discrete-time settings,  Recurrent Equilibrium Networks (RENs) have been proposed in~\cite{revay2023recurrent}, which enjoy built-in stability and robustness properties. Notably,  subsets of RENs can satisfy desired integral quadratic constraints by design and ensure a finite $\mathcal{L}_2$-gain. 
Despite their flexibility, RENs are computationally demanding, as their forward pass is not parallelizable in the length of the input sequence and, in general, requires solving an implicit equation.
% Moreover, RENs enforce either contractivity \cite{FB-CTDS} or incremental stability—a stronger form of robustness—by design, which can limit their expressiveness compared to general neural networks that only guarantee a finite $\mathcal{L}_2$ gain. 
% In contrast, our framework allows a more flexible choice of parameters since we allow for complex parameters. Moreover, parallel scan cannot be implemented for RENs making them slower to train as compared to SSMs.  
For a broader overview of NN-based approaches to control dynamical systems, we refer readers to~\cite{item_aba41bf93fe5484e9752d405af44acf6}.

\textbf{Contributions:} The main contributions of this paper are as follows:
\begin{itemize}
    \item We propose a novel free parametrization of the SSM-based L2RU architecture that enforces a prescribed $\mathcal{L}_2$-gain by design, without requiring computationally expensive projections, constrained optimization, or post-hoc verification. As an application, we combine this parametrization with the small-gain theorem to guarantee closed-loop stability both during training and at deployment.
    \item We demonstrate the effectiveness of the proposed approach through several experiments, including formation control of wheeled robots with collision avoidance among agents and with static obstacles.
\end{itemize}

\textbf{Organization:}
Section~\ref{sec:Prelims} introduces preliminaries on dynamical systems and SSMs and formulates the control problem.
Section~\ref{sec:Main_results} presents our main result: a free parametrization of the L2RU architecture with prescribed $\mathcal{L}_2$-gain.
Section~\ref{sec:experiments} reports numerical experiments illustrating the effectiveness of the proposed parametrization when used to parametrize a controller.
Finally, conclusions are drawn in Section~\ref{sec:conclusions}.

{\bf Notation:}
Throughout the paper, vectors are denoted by lowercase letters, matrices by uppercase letters, and sets by calligraphic letters. Sequences of vectors are denoted by bold lowercase letters. The set of all sequences $\mathbf{v} = (v_0,v_1,v_2,\dots)$ with $v_t \in \mathbb{R}^n$ for all $t \in \mathbb{N}$ is denoted by $\mathcal{L}^n$. We say that $\mathbf{v}$ belongs to the set of square-summable sequences $\mathcal{L}_2^n \subset \mathcal{L}^n$ if $\|\mathbf{v}\|_2 = \left( \sum_{t=0}^{\infty} \|v_t\|_2^2 \right)^{1/2} < \infty$.

A map (or operator) $g : \mathcal{L}^n \rightarrow \mathcal{L}^m$ is said to be causal if, for every $\mathbf{v} \in \mathcal{L}^n$, the output sequence $g(\mathbf{v}) = (w_0,w_1,\dots)$ satisfies $w_t = g_t(v_{0:t})$ for some maps $g_t$, where $v_{0:t} := (v_0,\dots,v_t)$ denotes the truncation of $\mathbf{v}$ to its first $t+1$ elements. For a causal map $g$, it is natural to define its action on finite sequences as $g(v_{0:t}) := g(\mathbf{v})_{0:t}$, where $\mathbf{v}$ is any extension of $v_{0:t}$ to an infinite sequence; causality ensures that this is well-defined.
With a slight abuse of notation, we use the same symbol $g(\mathbf{v})$ both for the full output sequence and for its truncation to a finite horizon whenever no ambiguity arises.
The expression $A \succ0$ ($A \succeq0$) defines a symmetric positive (semi) definite matrix $A$. 
% The spectrum (set of eigenvalues) of $A$ is denoted with $\lambda(A)$, its partition into conformal blocks with $A=\operatorname{Blk}(A_{11}, A_{12}, A_{21}, A_{22})$. 
The identity matrix is indicated with $I$, regardless of its dimension. $\mathds{1}_{n \times m}$ denotes a matrix of all ones with dimension $n \times m$. 

\section{Preliminaries and problem description}

\label{sec:Prelims}

\subsection{Dynamical systems}

We consider discrete-time input--output dynamical systems described in the state-space: 
\begin{equation} \label{eq:sys1}
    x_{k+1} = f(x_k,u_k), 
    \qquad 
    y_k = h(x_k, u_k),
    \qquad k \in \mathbb{N},
\end{equation}
with input $u_k \in \R^{n_u}$, output $y_k\in  \R^{n_y}$, and state $x_k \in \R^{n_x}$. 
We assume $f$ and $h$ are sufficiently well-behaved so that, for each initial condition $x_0 \in \mathbb{R}^{n_x}$ and input $\mathbf{u} \in \mathcal{L}^{n_u}$, unique trajectories $(\mathbf{x},\mathbf{y}) \in \mathcal{L}^{n_x}\times \mathcal{L}^{n_y}$ satisfies \eqref{eq:sys1}. This makes the transition map well-defined
\begin{equation}
    \varphi : \mathbb{R}^{n_x} \times \mathcal{L}^{n_u}
    \mapsto \mathcal{L}^{n_x} \times \mathcal{L}^{n_y}, 
    \qquad 
    (\mathbf{x},\mathbf{y}) = \varphi(x_0,\mathbf{u}),
\end{equation}
which makes the dependence on the initial condition explicit and characterizes the state and output trajectories of the system. For a fixed initial condition $x_0$, it is also useful to refer to the pure input/output map $\Sigma_{x_0} $:
\begin{equation}
   \Sigma_{x_0}  : \mathcal{L}^{n_u} \mapsto \mathcal{L}^{n_y}, 
    \qquad 
    \mathbf{y} = \Sigma_{x_0}(\mathbf{u}),
\end{equation}
which maps each input trajectory to the corresponding output trajectory, and it is simply obtained by projecting $\varphi$ onto the output component for a fixed $x_0$.
\footnote{More in general, $\Sigma_{\theta}$ can denote an input/output map depending on any parameter $\theta$, such as the initial condition. When the parameter is fixed, we shall drop it from the subscript for notational convenience.}
Note that the maps $\Sigma_{x_0}$ and $\varphi$ realized by a system in form \eqref{eq:sys1} are always causal. 
Finally, following a similar notation to \cite{van_der_schaft_small-gain_2017}, we use input/output maps to denote the standard closed-loop of two systems defined by the maps     $\Sigma_{x_0}^1 : \mc L_2^{n_u} \to \mc L_2^{n_y}$ and  $\Sigma_{h_0}^2 : \mc L_2^{n_y} \to \mc L_2^{n_u}$ as $\Sigma^1 ||_F \Sigma^2$ and we omit explicit reference to the initial conditions in this shorthand. We say that $\Sigma^1 ||_F \Sigma^2$ is well posed if for every $(x_0,h_0)$ the exist unique closed-loop trajectories $(\mathbf{y},\mathbf{u})$ satisfying
$
        \mathbf{y} = \Sigma_{x_0}^1(\mathbf{u}),
        \: \: 
        \mathbf{u} = \Sigma_{h_0}^2(\mathbf{y})
$.
% In the finite-horizon setting with horizon $N \in \mathbb{N}$, with a slight abuse of notation, we still use bold symbols to denote truncated trajectories, e.g.,
% \[
% \mathbf{u} = (u_0,\dots,u_{N-1}), \quad
% \mathbf{x} = (x_0,\dots,x_N), \quad
% \mathbf{y} = (y_0,\dots,y_N),
% \]
% and the operators $\Sigma_{x_0}$ and $\varphi$ are understood as their restrictions to these finite sequences.

In this work, we are interested in systems that possess a finite $\mc L_2$-gain (see \cite{van_der_schaft_small-gain_2017} for reference), for which we give the definition below.
\begin{defn}($\mc L_2$-\emph{gain}) \label{def:l2}
   Let $\Sigma_{x_0}: \mc L_2^{n_u} \mapsto \mc L_2^{n_y} $ be an input/output map associated with a dynamical system \eqref{eq:sys1} with initial condition $x_0 \in \R^{n_x}$. The system is said to have finite $\mc L_2$-gain if there exists $\gamma >0$ and a function $\beta(\cdot) : \R^{n_x} \mapsto \R^+$ such that, for any sequence $\mathbf{u} \in \mc L_2^{n_u}$ and $x_0 \in \R^{n_x}$ it holds
    \begin{equation}\label{eq:l2bound}
        \| \Sigma_{x_0}(\mathbf{u})\|_2 \le \gamma \|\mathbf{u}\|_2 +\beta(x_0) \: .  
    \end{equation}
  Any positive $\gamma$ such that \eqref{eq:l2bound} holds is called  an $\mc L_2$-bound of the system. The $\mc L_2$-gain is defined as the infimum among all $\gamma$ such that  \eqref{eq:l2bound} is satisfied for some function $\beta(\cdot)$.
\end{defn}
The concept of $\mc L_2$-bound is key for the well-known \emph{small-gain theorem}, a powerful tool in the stability analysis of possibly nonlinear systems connected in closed-loop.
\begin{thm}[Small-gain Theorem]
    Consider two causal input/output maps 
    $\Sigma_{x_0}^1 : \mc L_2^{n_u} \to \mc L_2^{n_y}$ and 
    $\Sigma_{h_0}^2 : \mc L_2^{n_y} \to \mc L_2^{n_u}$ realized by systems of the form 
    \eqref{eq:sys1} with initial conditions $x_0$ and $h_0$, respectively. 
    Assume that both maps satisfy \eqref{eq:l2bound} with $\mc L_2$-bounds $\gamma_1$ and $\gamma_2$ respectively and $\gamma_1 \gamma_2 <1$. Moreover, assume the closed-loop $\Sigma^1 ||_F \Sigma^2$ is well-posed. Then $\Sigma^1 ||_F \Sigma^2$ is $\mc L_2$-stable, in the sense that 
$ \mathbf{y} \in \mc L_2^{n_y},  \mathbf{u} \in \mc L_2^{n_u}$ for all $x_0 \in \R^{n_x}, h_0 \in \R^{n_h}$, where $
        \mathbf{y} = \Sigma_{x_0}^1(\mathbf{u}),
        \: \: 
        \mathbf{u} = \Sigma_{h_0}^2(\mathbf{y})
$ are the closed-loop trajectories. 
\label{thm:sg}
\end{thm}
% The small-gain theorem also holds under weaker hypotheses on the well-posedness of the closed loop, but the version stated here suffices for our purposes. 

\subsection{SSMs and the L2RU architecture}
\begin{figure}[t]
    \centering
\includegraphics[scale=0.78]{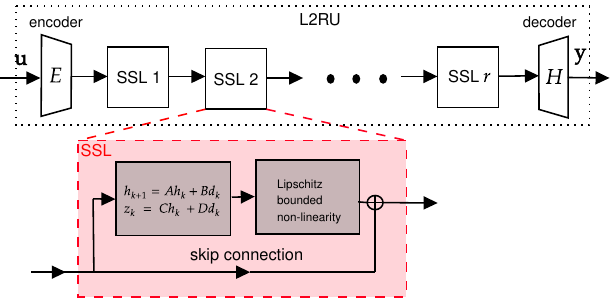}
    \caption{L2RU architecture. The model consists of a series of state-space layers, each comprised of $\mc L_2$-bounded DT LTI systems and Lipschitz-bounded nonlinearities. The input/output is pre- and post-processed by linear transformations.} 
    \label{fig:ssml2}
\end{figure}

In this section, we provide some preliminaries regarding SSMs.  While SSMs can take various forms, they are fundamentally characterized by a dynamical system consisting of multiple layers, each composed of LTI systems followed by a static nonlinear function.
Specifically, here we consider the L2RU architecture shown in Fig. \ref{fig:ssml2} and first introduced in \cite{massai_free_2025}, which resembles the Linear Recurrent Unit (LRU) \citep{orvieto_resurrecting_2023} but differs in the way the LTI system is parametrized, as we will see later on.
An L2RU is defined by the following components:
\begin{itemize}[leftmargin=10pt]
    \item \emph{State-space layer (SSL)}: a block consisting of a Discrete-Time (DT) LTI system followed by a static nonlinearity.  The LTI system is described in state space by:
    \begin{align} \label{eq:sysl}
g_{(A,B,C,D)} : \begin{cases}
        h_{k+1} &=Ah_k+Bd_k, \: \: h_0=0  \\
    z_k & = Ch_k+Dd_k \: ,
\end{cases}
\end{align}
where $h\in \R^{n_h} $ is the state, $d \in \R^{n_d},z \in \R^{n_z} $ are the input/output respectively and $A,B,C,D$ are the matrices, with appropriate dimensions, describing the dynamics. 
% The LTI system serves as the fundamental dynamical component of any SSM architecture, endowing the model with its memory capabilities. 
We denote the input/output map realized by \eqref{eq:sysl} as $\Sigma_{(A,B,C,D)}^l:\mc L^{n_d} \mapsto \mc L^{n_z}$ or simply as \(\Sigma^l\) when the explicit dependence on the system matrices can be omitted (note that the initial condition is fixed, so we drop it from the subscript).
%We focus on square systems primarily for technical reasons, as this enables an explicit, non-conservative parametrization of $\mc L_2$-bounded systems. This is not a major limitation since $n$ is a tunable hyperparameter, and in many SSM architectures, the LTI system's input and output naturally share the same dimension. Moreover, the input/output ($u$ and $y$ in Fig. \ref{fig:ssml2}) dimensions of the overall L2RU model are arbitrary and decoupled from $n$ via a linear encoder and decoder, as discussed later. 
% The L2RU architecture parametrizes the matrices $A, B, C, D$ such that the map $\Sigma_{(A,B,C,D)}^l$ satisfies \eqref{eq:l2bound} for a prescribed bound $\gamma$, or, equivalently, the associated system  $g_{(A,B,C,D)}$ has a prescribed finite $\mathcal{L}_2$-bound according to Definition \ref{def:l2}. In~\cite{massai_free_2025}, one such parametrization was first introduced. 
%In the LRU architecture, for instance, the authors exploit a similarity transformation in the complex field to parametrize a Schur-stable  $A$ as a complex diagonal matrix with eigenvalues that are always confined inside the unitary circle.
    The output of $\Sigma^l$ is fed to a static nonlinearity that belongs to a family of Lipschitz-bounded nonlinear functions 
    \begin{align} \label{eq:mu}
        \mu_{\xi} : \R^{n_z} \mapsto \R^{n_d} \:,
    \end{align}
    depending on the parameter $\xi \in \R^m$, with Lipschitz-bound in the $\mathcal{L}_2$-norm $\zeta$ for every value of $\xi$
    % \footnote{A function $\mu: \R^n \mapsto \R^m$ has 2-Lipschitz-bound $\zeta>0$ if $\|\mu(a)-\mu(b)\|_2 \le \zeta \|a-b\|_2$ for all $a,b \in \R^n$. For brevity, we refer to it as the Lipschitz-bound of $\mu$. } 
    and such that $\mu(0)=0$ (we omit the subscript when it is not necessary). Under this assumptions, it holds $\|\mu_\xi(\mathbf{u}) \|_2 \le \zeta  \|\mathbf{u} \|_2$ for all $\xi \in \R^m$ where $\mu(\mathbf{u})$ is the sequence obtained by applying $\mu$ to any input $\mathbf{u}$ element-wise. The parametrized family $\mu_\xi$ can range from simple functions to deep Lipschitz-bounded Multi-Layered Perceptrons (MLPs), such as the one proposed in \cite{wang_direct_2023}. A feed-through connection additively combines the input of the layer with its output. 
    % Such skip connections are widely used in deep learning to mitigate vanishing gradients and improve the information flow \cite{orhan_skip_2018}.  
    % % Finally, this type of block can be repeated and stacked in layers to form deep architectures and enhance the model expressivity. 
    \item \emph{Encoder/decoder}: 
   the input/output of the stacked SSLs are pre- and post-processed by linear encoder/decoder defined by the generic matrices $E \in \R^{n_d \times n_u}, H \in \R^{n_y \times n_z}$, where $n_u, n_y \in \mathbb{N}$ are the dimensions of input and output respectively. 
   % These are just linear transformations of the form $\mathbf{y}=E\mathbf{u}$ where $E$ multiplies the sequence $\mathbf{u}$ element-wise. 
\end{itemize}
For a fixed number of layers $r \in \mathbb{N}$, the construction defined above defines a dynamical system described by the input/output map $\Sigma_{\theta} : \mc L^{n_u} \mapsto \mc L^{n_y}$ uniquely determined by the following parameters:
\begin{align}\label{eq:setP}
    \theta \in \mc P&=\{ \{\underbrace{A_i,B_i,C_i,D_i}_{ \Sigma_i^l}, \underbrace{\xi_i}_{\mu_i}\}_{1 \le i \le r}, E, H \}  \;,
    % \notag \\
    % &=\R^{r(n_h^2+n_hn_d+n_zn_h+n_zn_d+m)+n_hn_u+n_zn_y} \: ,
\end{align} 
where $\Sigma_i^l$ and $\mu_i$ are the maps associated with the linear system and the nonlinear function of the generic $i$-th layer respectively.
The way $\Sigma_{\theta}$ acts on an input sequence $\mathbf{u}$ can be written more explicitly as a series composition of the encoder, $r$ SSLs and decoder:
\begin{align}\label{eq:l2ruseries}
    &\text{(encoder)} \quad & \mathbf{y}_0 &= E \mathbf{u}  \notag \\
    &\text{(SSL)} \quad & \mathbf{y}_i &= \mu_i\left( \Sigma_i^l ( \mathbf{y}_{i-1}) \right)+\mathbf{y}_{i-1}, \quad 1\le i \le r  \notag  \\
    &\text{(decoder)} \quad & \mathbf{y} &= H\mathbf{y}_{r} \: . 
\end{align}

The L2RU architecture leverages the parametrization of $\mathcal{L}_2$-bounded systems mentioned earlier to enforce a global $\mathcal{L}_2$-bound on the entire map $\Sigma_\theta$. More formally, in~\cite{massai_free_2025}, L2RU is defined through a \emph{free parametrization}, i.e., a piece-wise differentiable map $\rho_\gamma : \mathbb{R}^q \to \mathcal{P}$ such that $\Sigma_{\rho_\gamma(\omega)}$ has $\mathcal{L}_2\text{-bound } \gamma$ for all $\omega \in \mathbb{R}^q$, where $q \in \mathbb{N}$ denotes the number of free parameters. The parametrization is called \emph{free} because the parameter $\omega$ ranges over $\R^q$, therefore enabling unconstrained optimization with respect to it.
 For notational convenience, when no ambiguity arises, we will write $\Sigma_{\omega}^\gamma$ to denote an L2RU map, omitting the explicit parametrization $\rho$.

\subsection{Problem Formulation}

In this paper, we use the L2RU architecture to parametrize dynamical controllers
\[
\Sigma_{\omega}^{\gamma} : \mc L^{n} \mapsto \mc L^{m},
\]
with a prescribed finite $\mathcal{L}_2$-bound $\gamma$ for all $\omega \in \R^q$.

We employ such controllers in feedback with a possibly nonlinear $\mathcal{L}_2$-bounded discrete-time dynamical system, modeled as the causal map
\[
\varphi_s^{\hat{\gamma}}: \R^{n_x} \times \mc L^{n_u} \mapsto \mc L^{n_x} \times \mc L^{n_y}, 
\quad (x_{0_s}, \mathbf{u}) \mapsto (\mathbf{x}_s,\mathbf{y}_s),
\]
where $s$ indexes a particular realization of the initial state $x_{0_s}$ (and, in turn, of the resulting trajectories $\mathbf{x}_s$ and $\mathbf{y}_s$). We assume that the plant has a known finite $\mathcal{L}_2$-bound $\hat{\gamma}$.

Our goal is to choose the controller parameters $\omega$ so as to stabilize the closed-loop interconnection while minimizing a general almost-everywhere differentiable objective function
\[
c : \mc L^{n_x} \times \mc L^{n_y} \times \mc L^{n_u} \mapsto \R,
\quad (\mathbf{x},\mathbf{y},\mathbf{u}) \mapsto c(\mathbf{x},\mathbf{y},\mathbf{u}).
\]
In practice, we approximate the expected performance by an empirical average over $S$ samples of the initial condition:
% \footnote{The empirical average in the cost is a Monte Carlo approximation of the exact, and often intractable, expectation with respect to a probability distribution over initial conditions.}
\begin{align} \label{Chap2:opt_prob}
        \displaystyle\min_{{\omega \in \R^q}} \quad &  
        \frac{1}{S} \sum_{s=1}^S c(\mathbf{x}_s^{\omega}, \mathbf{y}_s^{\omega}, \mathbf{u}_s^{\omega})  \notag  \\
        \text{s.t.} \quad 
        & (\mathbf{x}_s^{\omega}, \mathbf{y}_s^{\omega}) 
          = \varphi_s^{\hat{\gamma}}(x_{0_s}, \mathbf{u}_s^{\omega}) 
          \quad && \text{(system dynamics)} \notag \\ 
        & \mathbf{u}_s^{\omega} 
          = \Sigma_{\omega}^{\gamma}(\mathbf{x}_s^{\omega}) 
          \quad && \text{(L2RU controller).}
\end{align}
We make the standard assumption that the feedback interconnection is well-posed. That is, for every initial state $x_{0_s}$ and every $\omega \in \R^q$, there exists a unique closed-loop trajectory $(\mathbf{u}_s^{\omega}, \mathbf{x}_s^{\omega}, \mathbf{y}_s^{\omega})$ consistent with the system and L2RU dynamics in problem~\eqref{Chap2:opt_prob}. In practice, we are often interested in finite-horizon optimal control problems, where the cost can be written as a sum of stage costs $\ell(\cdot)$ plus a terminal cost $\phi(\cdot)$ over an horizon $T \in \mathbb{N}$:
\begin{equation} \label{Chap2:eq:loss_function}
     c(x_{0:T}, y_{0:T}, u_{0:T}) 
     = \frac{1}{T} \sum_{k = 0}^{T-1} \ell(x_k, y_k, u_k) + \phi(x_T) \, .
\end{equation}
The formulation in~\eqref{Chap2:opt_prob} encompasses this case as well. As explained in the notation section, the causal maps $\varphi$ and $\Sigma$ are then understood as their restriction to the finite horizon $T$.

We aim to train an L2RU controller with guaranteed $\mathcal{L}_2$-bound $\gamma$ and then for the system $\varphi_s^{\hat{\gamma}}$ with $\mathcal{L}_2$-bound $\hat{\gamma}$, the small-gain theorem ensures $\mathcal{L}_2$-stability of the closed-loop interconnection whenever $\gamma \hat{\gamma} < 1$ according to Theorem \ref{thm:sg}. In particular, this can be enforced by setting
$
\gamma = \frac{1}{\hat{\gamma} + \varepsilon},
$
for some arbitrarily small $\varepsilon > 0$.
To this end, we want to construct a {free parametrization} of the L2RU architecture, that is, a map
$
\rho_\gamma : \R^q \to \mc P
$
such that, for every $\omega \in \R^q$, the corresponding controller $\Sigma_{\rho_\gamma(\omega)}^{\gamma}$ has $\mathcal{L}_2$-bound $\gamma$.

Moreover, we seek a parametrization that enables parallelization via parallel scan algorithms by enforcing a structured form of the matrix $A$, akin to the complex-diagonal form used in the original LRU architecture~\cite{orvieto_resurrecting_2023}. This structure unlocks significant efficiency gains when solving the Optimal Control Problem (OCP)~\eqref{Chap2:opt_prob} whenever the time recursion that realizes the plant dynamics $\varphi_s^{\hat{\gamma}}$ is also parallelizable. This is, in particular, the case when the plant is linear, or when it is itself described by an SSM architecture whose dynamical component admits a scan implementation, such as the LRU.

\section{Main Results}
\label{sec:Main_results}
In the following, we present a novel free parametrization of $\mc L_2$-bounded LTI systems with bound ${\gamma}$ that generalizes the one originally introduced in \cite{massai_free_2025}.

\begin{thm}
 $\kappa_{\gamma}$ defined in Block \ref{box3} is a free parametrization of $\mc L_2$-bounded LTI systems with bound ${\gamma}$.
\label{thm:param_suff}
\end{thm}
The proof of Theorem~\ref{thm:param_suff} is provided in the accompanying technical report \cite{massai_l2ru:_2025}. In essence, it shows how the parameters can be freely chosen while satisfying the bounded real lemma.
The map $\kappa_{\gamma}$ parametrizes LTI systems with arbitrary dimensions $n_h,n_u,n_y$ and thus generalizes the construction in~\cite{massai_free_2025}, which is restricted to square systems with $n_h = n_u = n_y$. However, $\kappa_{\gamma}$ is not a \emph{complete} parametrization of $\mc L_2$-bounded LTI systems with bound $\gamma$, since its construction relies on sufficient, but not necessary, conditions, as detailed in the proof of Theorem~\ref{thm:param_suff}. Despite this, $\kappa_{\gamma}$ can produce systems whose $\mc L_2$-gain is very close to the prescribed bound $\gamma$, particularly for small values of $\epsilon$ in~\eqref{eq:D}.

The parametrization is motivated by the optimal control problem~\eqref{Chap2:opt_prob}: once $\gamma$ is fixed to satisfy the small-gain condition with the plant, closed-loop stability is guaranteed by design, and the optimization over $\omega$ can be carried out as an unconstrained nonlinear optimization problem that focuses purely on performance and can be efficiently solved via gradient descent and backpropagation in time. As mentioned in the previous section, the parametrization $\rho_{\gamma}$ is constructed by first deriving a parametrization of $\mathcal{L}_2$-bounded linear systems of the form~\eqref{eq:sysl}. The key problem we address is therefore that of finding,  given $\tilde{\gamma}>0$ \footnote{We denote the bound by $\tilde{\gamma}$ just to avoid confusion with the $\mc L_2$-bound $\gamma$ of the whole L2RU map.},  a function $\kappa_{\tilde{\gamma}} : \mathbb{R}^p \mapsto \{ A,B,C,D\} $ such that $\Sigma_{\kappa_{\tilde{\gamma}} (\omega)}^l$ satisfies \eqref{eq:l2bound} for the prescribed bound $\tilde{\gamma}$ for all $\omega \in \R^p$, where $p \in \mathbb{N}$ is the number of free parameters. We call any such function a \emph{free parametrization of $\mc L_2$-bounded LTI systems with bound $\tilde{\gamma}$}.

\begin{mybox}{Parametrization of DT LTI systems with prescribed $\mc L_2$-gain $\gamma$}{box3}
Given $\gamma>0$ and the set of free parameters 
\begin{align*} \resizebox{.99\hsize}{!}{$ 
 \mc F = \left\{ (\{\mu_j, \theta_j\}_{j \in \{1, \ldots, n_h\}}, \tilde{D}, \bar{Y} ) \in \mathbb{R}^{2n_h(1 + (n_d + n_z)) + (n_d \times n_{z})} \right\}.$}
\end{align*}
% where
% \begin{align*} \mu_j \sim \mathcal{U}\!\left[ \log(-\log(\bar{r})), \, \log(-\log(r)) \right], \\ \theta_j \sim \mathcal{U}\!\left[ \log(\log(\theta)), \, \log(\log(\bar{\theta})) \right], \end{align*}
% with $0 \le r < \bar{r} < 1$ denoting the minimum and maximum modulus of each eigenvalue, respectively, and $0 \le \theta < \bar{\theta} < \pi$ denoting the corresponding minimum and maximum phase. Here, $\mathcal{U}[a,b]$ indicates the uniform distribution over the interval $[a,b]$.
Define $\kappa_{\gamma}: \omega \in  \mc F \mapsto (A, B, C, D)$ as follows \begin{align}  
& \lambda_j = \exp\!\left(-\exp(\mu_j) + i \exp(\theta_j)\right).  \label{eq:A1} \\
& A = \text{diag}(\lambda_1, \dots \lambda_{n_h}) \label{eq:A2} \\
& B = P^{-1}Y_{21}, \ 
 C = Y_{22}^\top,  \,
 D =  \frac{\gamma}{\|\tilde{D}\|_2 + \epsilon} \tilde{D} \label{eq:D}
\end{align}
where $Y$ and $P$ are defined as a function of the free parameters as follows:
    \begin{align*} 
   &Y = \eta^{-1} \tilde{Y}, \
    \tilde{Y} = \mathcal{M} \odot \bar{Y}, \ \mathcal{M} = \begin{bmatrix}
       \mathds{1}_{n \times n_u} & 0_{n \times n_y} \\ 0_{n \times n_u} & \mathds{1}_{n \times n_y}
   \end{bmatrix} \\
 & \eta = \max \{ 1, \max \{W^{-1} \tilde{Y}, Z^{-1} \tilde{Y} \} \} \\
 & Z = \begin{bmatrix}
     \gamma I & D^\top \\ D & \gamma I 
 \end{bmatrix}, \ W = \begin{bmatrix}
     P & P A \\ A^\top P & P
 \end{bmatrix}, \ P = A^\top A + \epsilon I_{n_h}.
\end{align*} 
\end{mybox}

\subsection*{Computational efficiency and initialization}
It is worth highlighting that the parametrization $\kappa_{\gamma}$ adopts the complex-diagonal parametrization of $A$ (see~\eqref{eq:A1}, \eqref{eq:A2}), as in the LRU architecture~\citep{orvieto_resurrecting_2023}. This structure enables highly efficient time-domain simulation via parallel scan algorithms (\cite{blelloch_prefix_2004}), substantially reducing trajectory generation time compared to standard recursive schemes. Moreover, since the complex eigenvalues of $A$ are parametrized directly, they can be initialized with moduli arbitrarily close to the boundary of the unit disk, which is particularly beneficial for long-memory behavior in L2RU models. In contrast, the parametrization in~\cite{massai_free_2025} yields a dense real matrix $A$, resulting in more costly simulation and a less effective initialization procedure. A more detailed discussion and a comparison with the original L2RU parametrization are provided in the extended report~\cite{massai_l2ru:_2025}.

\begin{rem}
The closed-loop stability guarantee based on the small-gain theorem hinges on our ability to compute or estimate an $\mc L_2$-bound of the plant, i.e., an upper bound on its actual $\mc L_2$-gain. This can be challenging for general nonlinear dynamical systems. Nevertheless, there is a broad class of systems for which a reasonable bound can be obtained, making our approach practically relevant. In particular, for Schur-stable linear systems and SSMs such as L2RU and LRU, an exact $\mc L_2$-bound can be computed. In purely data-driven settings, one typically obtains lower-bound-type estimates from finite simulations, which do not suffice for a formal small-gain certificate. Nevertheless, the proposed L2RU parametrization allows us to prescribe and tune the $\mathcal{L}_2$-bound of the controller.
\end{rem}

Finally, one can use $\kappa_{\tilde{\gamma}}$ to induce a free parametrization $\rho_{\gamma}$ of the L2RU with prescribed $\mc L_2$-bound $\gamma$, we refer the reader to \cite{massai_l2ru:_2025} for more details on parametrization. 
% \begin{mybox}{L2RUs parametrization}{box2}
% Given $\mu_{\xi}$ as defined in \eqref{eq:mu}, ${\gamma}>0$ and the set of free parameters
% \begin{align*}
%    \left\{ \{\omega_i,\tilde{\gamma}_i, \xi_i, \tilde{\zeta}_i\}_{1\le i\le r}, \tilde{E}, \tilde{H} \right\} = \R^{q}, 
% \end{align*}
% define the map $\rho_{{\gamma}} : \R^{q} \mapsto \mc P $, where $\mc P$ is given in \eqref{eq:setP}, as follows: 
% \begin{align}
% & (A_i, B_i, C_i, D_i) = \kappa_{\tilde{\gamma_i}}(\omega_i) \:, \:  \: 1\le i \le r \label{eq:psys}\\
% & \xi_i = \tilde{\xi_i} \:, \:  \: 1\le i \le r \\
% & E={\tilde{E}} \label{eq:pE} \\
% & H= \frac{\tilde{H}{\gamma}}{\|\tilde{H}\|_2 \|\tilde{E}\|_2} \displaystyle\prod_{i=1}^{r} \left( |\tilde{\gamma}_i\tilde{\zeta}_i| +1\right)^{-1} \label{eq:pF}
% \end{align}
% \end{mybox}
% The map $\rho_{\gamma}$ defined in Block~\ref{box2} is such that $\Sigma_{\rho_{\gamma}(\omega)}$ satisfies~\eqref{eq:l2bound} with bound $\gamma$, thereby parametrizing the L2RU architecture, as proved in~\cite{massai_free_2025} (Theorem~3).
It is worth noting that the $\mc L_2$-bounds and Lipschitz constants of the linear systems and nonlinearities in each layer are treated as free parameters in the definition of $\rho_{\gamma}$ \footnote{Technically $\gamma, \zeta$ are not free parameters as they must be positive, however, we can simply set $\gamma=|\tilde{\gamma}|, \zeta=|\tilde{\zeta}|$ where $\tilde{\gamma}, \tilde{\zeta}$ are free parameters (or, equivalently, use any smooth positive reparametrization).}. Their values need not be fixed a priori; instead, they are used to properly normalize the $\mc L_2$-bound of the decoder $H$, to guarantee that the overall bound of the L2RU architecture equals the prescribed $\gamma$. 
% The key point is that this requires explicitly computing the $\mc L_2$-bounds and Lipschitz constants, which is made possible by the parametrization $\kappa_{\gamma}$ for the linear systems and by the definition of $\mu_{\xi}$ for the nonlinearities.

\section{Simulation studies}
\label{sec:experiments}
In this section, we present an experiment illustrating the effectiveness of the proposed approach. We consider a swarm of robots that must navigate through the same environment while also avoiding mutual collisions. An additional experiment is provided in \cite{massai_l2ru:_2025}. The Python implementation of the L2RU architecture can be found at: \url{https://github.com/DecodEPFL/SSMs-for-Control/tree/master}.

% \subsection{Navigation of a Unicycle Robot Through Obstacles}
% \label{subsec:unicycle_navigation}

We consider a discretized nonlinear wheeled robot from the Robotarium platform \citep{wilson2020robotarium} as follows
% The continuous-time dynamics of the robot is given by:
% \[
% \dot{x} = \begin{bmatrix}
%     \cos(\theta) & 0 \\
%     \sin(\theta) & 0 \\
%     0 & 1
% \end{bmatrix}
% \begin{bmatrix}
%     v \\ \omega
% \end{bmatrix}, \; y = \begin{bmatrix}
%     1 & 0 & 0 \\
%     0 & 1 & 0
% \end{bmatrix} x,
% \]
\[
\begin{aligned}
x_{k+1} &= f(x_k, u_k) = x_k + \tau \begin{bmatrix}
    \cos(\theta_k) & 0 \\
    \sin(\theta_k) & 0 \\
    0 & 1
\end{bmatrix}
\begin{bmatrix}
    v_k \\ \omega_k
\end{bmatrix} = x_k + B_x u_k\\
y_k &= h(x_k, u_k) = Cx_k.
\end{aligned}
\]
where \( x^\top = [x_p, y_p, \theta]^\top \) represents the robot's state, with \((x_p, y_p)\) denoting its position and \(\theta\) its orientation with respect to the \(x\)-axis. The control inputs \( u^\top = [v, \omega]^\top \) correspond to the linear and angular velocities, respectively. 
To compute the \(\mathcal{L}_2\)-gain of the system, we employ the results from \cite[Theorem 3]{koelewijn2021incremental}, which require solving the following semi-definite program; see \cite{massai_l2ru:_2025} for more details.

% \[
% \begin{bmatrix}
% P & \mathcal{A}_{\delta}(x, u)P & \mathcal{B}_{\delta}(x,u) & 0 \\
% \star & P & 0 & P\mathcal{C}_{\delta}^{\top}(x,u) \\
% \star & \star & \gamma I & \mathcal{D}_{\delta}^{\top}(x,u) \\
% \star & \star & \star & \gamma I
% \end{bmatrix} \succ 0,
% \]
% where
% \[
% \mathcal{A}_{\delta} = \frac{\partial f}{\partial x}, \quad
% \mathcal{B}_{\delta} = \frac{\partial f}{\partial u}, \quad
% \mathcal{C}_{\delta} = \frac{\partial h}{\partial x}, \quad
% \mathcal{D}_{\delta} = \frac{\partial h}{\partial u},
% \]
% and \(P \succ 0\) is the decision variable.
% For this system, the Jacobians are computed as:
% \[
% \begin{aligned}
% \mathcal{A}_{\delta}(x,u) &= I + \tau \begin{bmatrix}
%     0 & 0 & -\sin(\theta_k) v_k \\
%     0 & 0 & \cos(\theta_k) v_k \\
%     0 & 0 & 0
% \end{bmatrix}, \\
% \mathcal{B}_{\delta}(x, u) &= \begin{bmatrix}
%     \cos(\theta_k) & 0 \\
%     \sin(\theta_k) & 0 \\
%     0 & 1
% \end{bmatrix}, \
% \mathcal{C}_{\delta}(x,u) = C, \
% \mathcal{D}_{\delta}(x,u) = 0.
% \end{aligned}
% \]

% We define the parameter vector \(\rho = [\rho_1, \rho_2, \rho_3, \rho_4]^\top\), where:
% \[
% \begin{aligned}
% \rho_1 &= -\sin(\theta_k) v_k \in [-v_{max}, v_{max}], \\
% \rho_2 &= \cos(\theta_k) v_k \in [-v_{max}, v_{max}], \\
% \rho_3 &= \cos(\theta_k) \in [-1, 1], \
% \rho_4 = \sin(\theta_k) \in [-1, 1].
% \end{aligned}
% \]
% This results in \(2^4 = 16\) linear matrix inequalities (LMIs) on the vertices of \(\rho\) to compute the \(\mathcal{L}_2\)-gain, which, using the CVXPY package, 
The $\mathcal{L_2}$ is found to be \(18.573\). We then train an L2RU controller with \(\mathcal{L}_2\)-bound smaller than \(1/18.573\). We parametrize the controller with an L2RU with $r=3$ layers and dimensions $
n_d = n_z = 12, n_h = 11 $  and $ \gamma = 1/(18.57+0.001)$. For the nonlinearities $\mu_\xi$, we employed the Lipschitz-bounded MLPs presented in \cite{wang_direct_2023}.
We choose the learning rate of $\texttt{1e-3}$ with a linear scheduler and Adam as the optimizer.

% The controller is trained for 10000 epochs to minimize the following cost function:
% \[
% \ell(\theta) = \frac{1}{TS} \sum_{s=1}^{20} \left( \sum_{k=1}^T \lambda_{goal} \ell_{goal} + \lambda_{u} \ell_{u} + \lambda_{obs} \ell_{obs} \right),
% \]
% where \(\lambda_{goal}\), \(\lambda_{u}\), and \(\lambda_{ca}\) are regularization parameters for the cost terms:
% \[
% \begin{aligned}
% \ell_{goal} &= \|x^{star} - x_k\|,  \quad \quad  \ell_{u} = u_k^\top R u_k,\\
% \ell_{obs} &= \frac{1}{N_{obs}} \sum_{i=1}^{N_{obs}} \left[ \max \left( 0, (r_i + d_{safe}) - \|x_t - c_i\| \right) \right]^2.
% \end{aligned}
% \]

% The closed-loop simulation results are shown in Fig.~\ref{fig:single_agent}, demonstrating successful navigation from start to goal.

% \begin{figure}[htbp]
%     \centering
%     \includegraphics[scale=.45]{figures/unicycle_trajectory.pdf}
%     \caption{Trained L2RU controller navigating the robot through obstacles.}
%     \label{fig:single_agent}
% \end{figure}

% \subsection{Formation Control Subject to Collision Avoidance}
% \label{subsec:formation_control}

% In this experiment, we extend the previous setup to a homogeneous swarm of wheeled robots, solving a multi-agent formation control problem while avoiding collisions with obstacles and among agents. 
We assume a centralized controller that receives information from all robots and provides control inputs. The dynamics of the swarm are given by:
\[
{x}_{k+1} = I_N \otimes _k + (I_N \otimes B_x) {u}_k,
\]
where \(N\) is the number of agents, \({x}_k^\top = [x_{1,k}^\top, x_{2,k}^\top, \dots, x_{N,k}^\top]\), and \(\bm{u}_k^\top = [u_{1,k}^\top, u_{2,k}^\top, \dots, u_{N,k}^\top]\). The \(\mathcal{L}_2\)-gain of the swarm system is identical to that of individual agents, as the dynamics are decoupled \citep{arcak2018networks}.

\begin{figure}
    \centering
    \includegraphics[scale=.42]{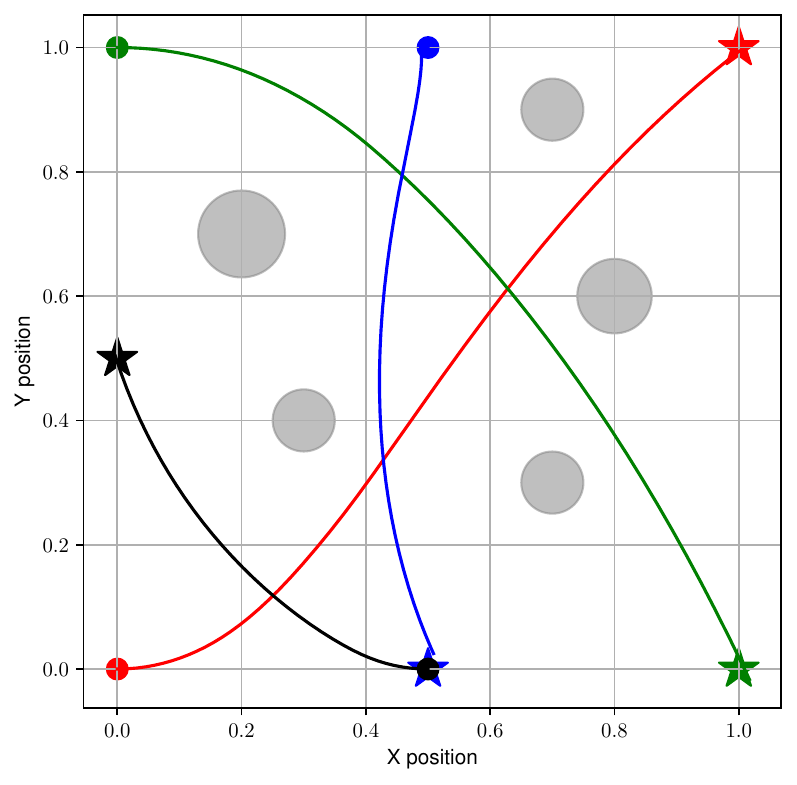}
\caption{Trajectories of four agents from their initial positions to their goals while avoiding both obstacles and mutual collisions. Gray circles represent obstacles, solid circles denote starting positions, and stars indicate goal positions.}
    \label{fig:multi-agent-trajectories}
\end{figure}

\begin{figure}
    \centering
    \includegraphics[scale=.45]{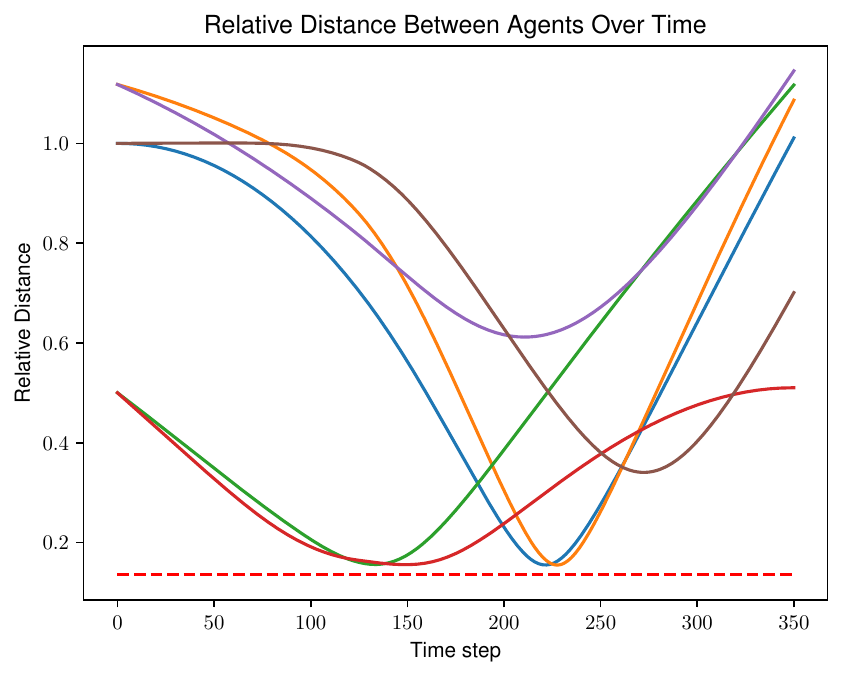}
\caption{Relative distances between agents over time, showing that no collisions occur. }
    \label{fig:relative_distances}
\end{figure}

The centralized L2RU controller has dimensions \(n_{{u}}^c = N \times n_u\) and \(n_{{y}}^c = N \times n_y\). The cost function to minimize is:
\[
\resizebox{\columnwidth}{!}{$
\ell(\theta)=\frac{1}{TS}\sum_{s=1}^{20}\left(\sum_{k=1}^T
\lambda_{goal}\ell_{goal}
+\lambda_u\ell_u
+\lambda_{obs}\ell_{obs}
+\lambda_{ca}\ell_{ca}\right)
$}
\]
where the new cost term is defined as:
\[
\begin{aligned}
\ell_{goal} &= \|x^{star} - x_k\|, \
\ell_{u} = u_k^\top R u_k, \\
\ell_{obs} &= \frac{1}{N_{obs}} \sum_{i=1}^{N_{obs}} \left[ \max \left( 0, (r_i + d_{safe}) - \|x_t - c_i\| \right) \right]^2, \\
\ell_{ca} &= \begin{cases}
\frac{1}{N^2} \sum_{i=1}^{N-1} \sum_{j=i+1}^N \frac{1}{(r_{ij} + \epsilon)^2}, & \text{if } r_{ij} < \bar{r}, \\
0, & \text{otherwise},
\end{cases}
\end{aligned}
\]
with \(r_{ij} = \sqrt{(x_{i,k} - x_{j,k})^2 + (y_{i,k} - y_{j,k})^2}\) as the relative distance between agents \(i\) and \(j\). Here, \(N_{obs}\) is the number of obstacles, \(d_{safe} = d_{margin} + r_{obs} + r_{robot}\) is the safe distance, and \(c_i\) is the center of the \(i\)-th obstacle. The collision-avoidance distance \(\bar{r}\) is set to \(0.185\,\text{m}\), ensuring no collisions for \(r_{ij} < 0.165\,\text{m}\). The results are provided in Figures \ref{fig:multi-agent-trajectories} and \ref{fig:relative_distances}. Both plots demonstrate that the objectives are successfully achieved, and constraining the controller's $\mc L_2$-bound to $1/18.573$ does not result in any appreciable loss of expressivity.

\section{Conclusion}
\label{sec:conclusions}

The control of nonlinear systems poses several challenges, chief among them ensuring closed-loop stability while maintaining computational tractability. 
To address this issue, we have proposed a novel free parametrization of the L2RU, an SSM architecture that preserves closed-loop stability by enforcing a prescribed finite $\mathcal{L}_2$-bound and leveraging the small gain theorem. 
Moreover, since the proposed controllers are designed in discrete time and their forward passes are highly parallelizable, the implementation of the trained policies can be significantly more efficient than that of NODE-based controllers. 
Future work will focus on deploying the controllers on real-world systems and on exploiting this parametrization within model-based Reinforcement Learning (RL) frameworks and considering the weighted $\mathcal{L}_2$ norm.

% \begin{ack}
% Place acknowledgments here.
% \end{ack}

% \section*{DECLARATION OF GENERATIVE AI AND AI-ASSISTED TECHNOLOGIES IN THE WRITING PROCESS}
% During the preparation of this work the author(s) used [NAME TOOL / SERVICE] in order to [REASON]. After using this tool/service, the author(s) reviewed and edited the content as needed and take(s) full responsibility for the content of the publication.

\balance
\bibliography{ifacconf}             % bib file to produce the bibliography
                                                     % with bibtex (preferred)
                                                   
%\begin{thebibliography}{xx}  % you can also add the bibliography by hand

%\bibitem[Able(1956)]{Abl:56}
%B.C. Able.
%\newblock Nucleic acid content of microscope.
%\newblock \emph{Nature}, 135:\penalty0 7--9, 1956.

%\bibitem[Able et~al.(1954)Able, Tagg, and Rush]{AbTaRu:54}
%B.C. Able, R.A. Tagg, and M.~Rush.
%\newblock Enzyme-catalyzed cellular transanimations.
%\newblock In A.F. Round, editor, \emph{Advances in Enzymology}, volume~2, pages
%  125--247. Academic Press, New York, 3rd edition, 1954.

%\bibitem[Keohane(1958)]{Keo:58}
%\newblock \emph{Power and Interdependence: World Politics in Transitions}.
%\newblock Little, Brown \& Co., Boston, 1958.

%\bibitem[Powers(1985)]{Pow:85}
%\newblock Is there a way out?
%\newblock \emph{Harpers}, pages 35--47, June 1985.

%\bibitem[Soukhanov(1992)]{Heritage:92}
%\newblock \emph{{The American Heritage. Dictionary of the American Language}}.
%\newblock Houghton Mifflin Company, 1992.

%\end{thebibliography}

% \appendix
% \section{A summary of Latin grammar}    % Each appendix must have a short title.
% \section{Some Latin vocabulary}              % Sections and subsections are supported  
%                                                                          % in the appendices.
\end{document}